\documentclass{elsart}

\usepackage{epsfig}

\usepackage{amssymb}

\begin{document}

\begin{frontmatter}



\title{Direct Photon Production in Au+Au Collisions at RHIC-PHENIX Experiment}


\author[CNS]{Tadaaki~Isobe\thanksref{TU}} for the PHENIX collaboration
\address[CNS]{Center for Nuclear Study, Graduate School of Science,
 University of Tokyo, 7-3-1 Hongo, Bunkyo, Tokyo 113-0033, Japan}
\thanks[TU]{isobe@cns.s.u-tokyo.ac.jp, Tel.:+81 48 464 4156, Fax:+81 48
 464 4554}

\begin{abstract}
 Direct photons have been measured with the PHENIX experiment in Au+Au
 collisions at $\sqrt{s_\mathrm{NN}}$ = 200~GeV. 
 The direct photon result obtained with PHENIX-EMCal up to 18~GeV/$c$
 is consistent with the NLO pQCD calculation scaled by the nuclear
 overlap function. 
 The measurement using internal conversion of photons into $e^+e^-$
 shows the enhancement of the yield comparing with NLO pQCD calculation.
\end{abstract}

\end{frontmatter}

% main text
\section{Introduction}
\label{intro}

 Since photons do not interact strongly, direct photons are a powerful
 probe to study the initial state of matter produced in relativistic
 heavy ion collisions.  
 They are emitted from all the states such as the initial state, the
 Quark-Gluon Plasma (QGP), and the final hadron-gas state.    
 In addition, it is predicted that the contribution of photons from
 jet-photon conversion in the dense medium can be as large as the photon
 yield from hard scatterings~\cite{bib:fries}.

 In the 2004 Run, PHENIX recorded a high-statistics
 Au+Au data set.
 The new data set allows us to measure direct photons beyond the
 transverse momentum ($p_\mathrm{T}$) of 10~GeV/$c$  and also more
 precisely than before at intermediate $p_\mathrm{T}$ where thermal
 photons and photons from jet-plasma interactions are important. 
 The PHENIX experiment can measure photons with two types of highly
 segmented electromagnetic calorimeters~(EMCal)~\cite{bib:emcal}. 
 One is a lead scintillator sampling calorimeter~(PbSc), and another is
 a lead glass Cherenkov calorimeter~(PbGl).

\section{Measurement of mid-$p_\mathrm{T}$ direct photon}
\label{soft}

Measurement of direct photons is challenging because there is a
large amount of background from decay of neutral mesons such as $\pi^0$
and $\eta$.
It is estimated that the
$\gamma/\gamma_{bg}$ would be $\sim$ 10~\% at the $p_\mathrm{T}$ range
of 2-4~GeV/$c$, where thermal photon is expected to be
dominant~\cite{bib:thermalphoton}. 
In order to extract direct photon signal, conventional subtraction
method has been used.
$\pi^0$ and $\eta$ mesons are reconstructed via their two-photon decay
mode. 
The $p_\mathrm{T}$ spectra of direct photons are obtained by subtracting
the spectra of decay photons estimated based on the measured
$\pi^0$/$\eta$ from the $p_\mathrm{T}$ spectra of inclusive photons.

\begin{figure}[h]
 \begin{center}
  \includegraphics*[width=5.6cm]{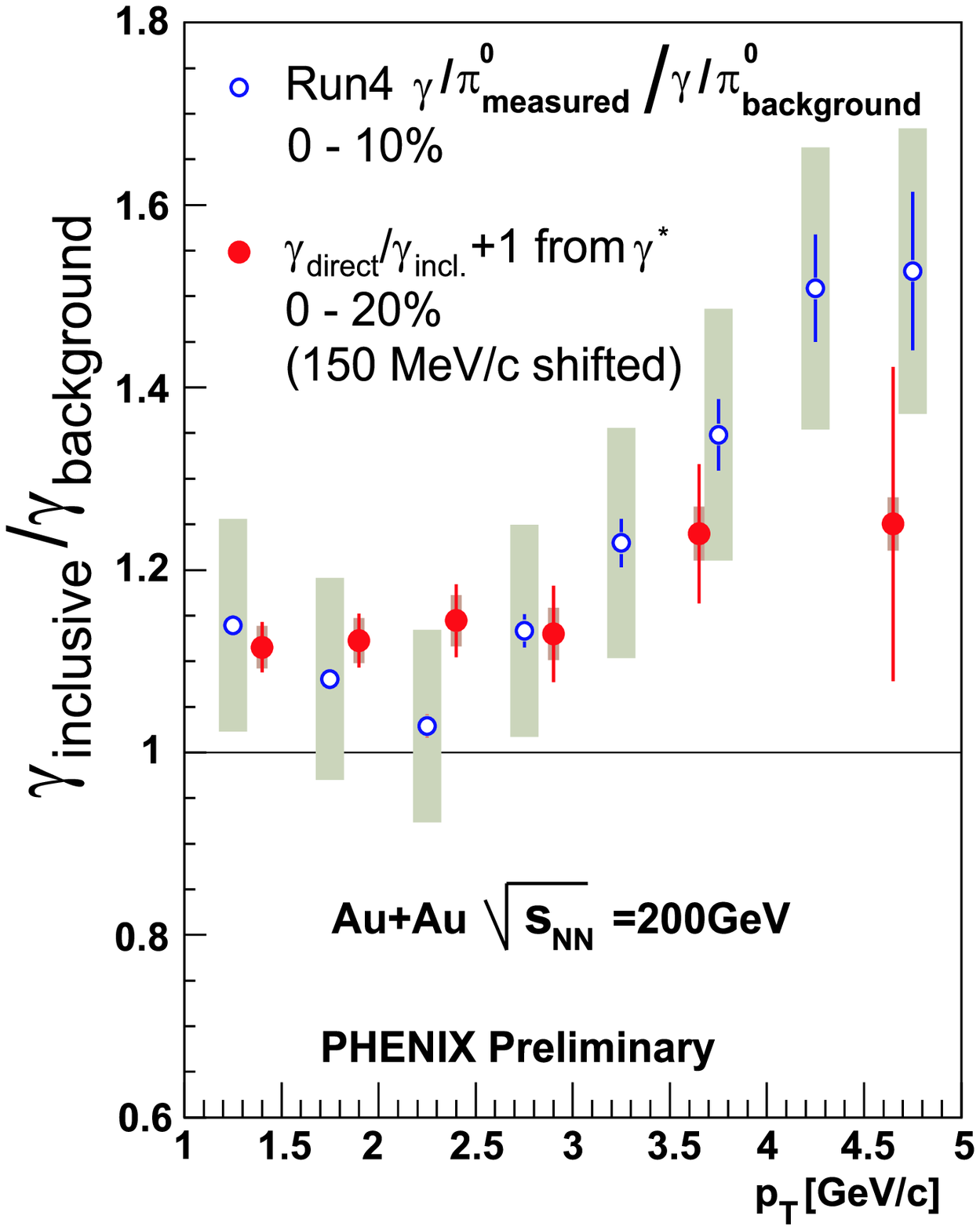}
  \includegraphics*[width=5.6cm]{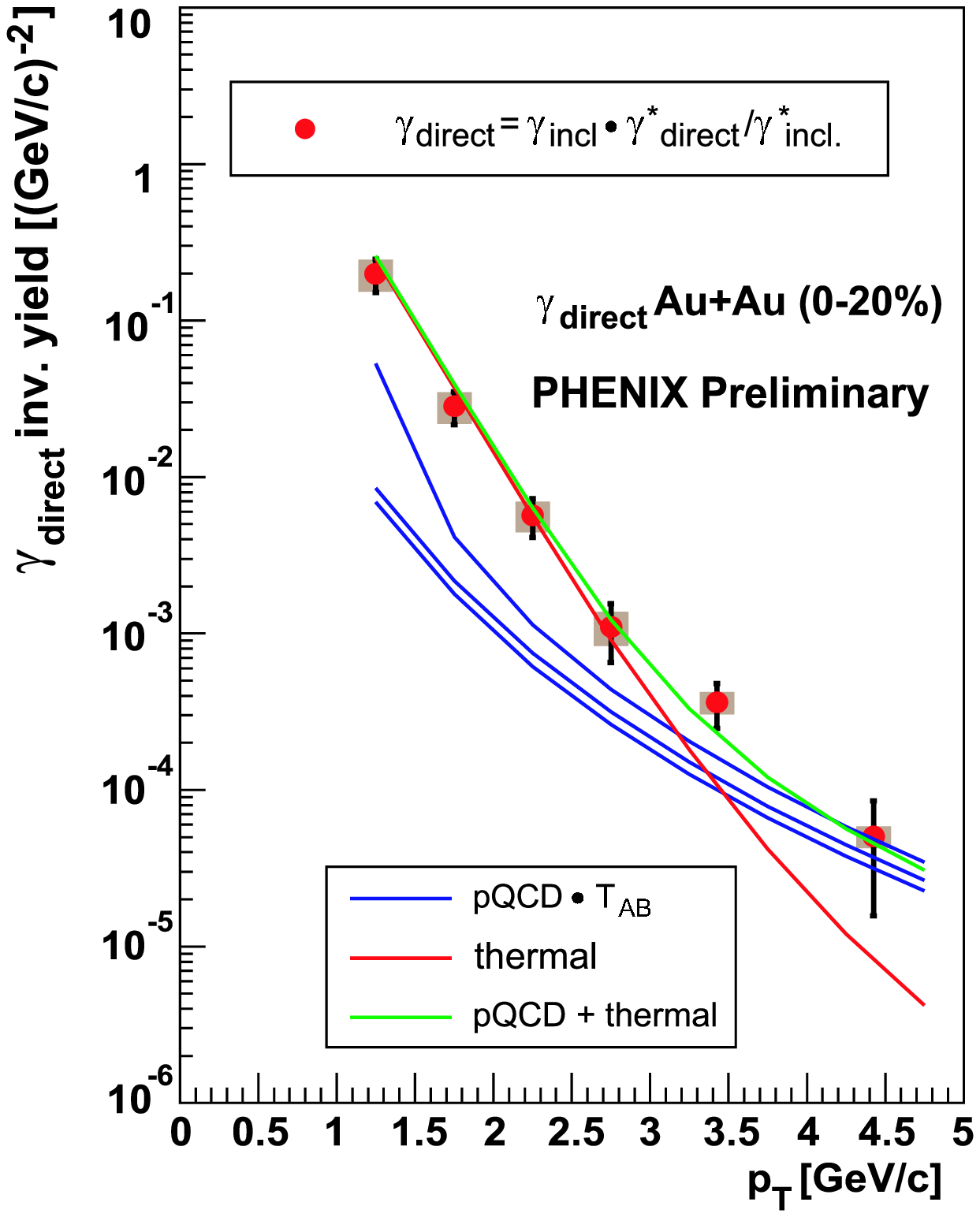}
 \end{center}
% \vspace{-6mm}  % to adjust space between the figure and caption
 \caption{Left: Direct photon excess ratio measured with subtraction
 method and with internal conversion into di-electron., Right: The
 direct photon spectrum from the measurement of internal conversion
 compared to NLO pQCD and hydrodynamics (QGP with $\langle T_0 \rangle$
 = 360~MeV) predictions.} 
 \label{fig:midpt}
\end{figure}

In addition to the conventional subtraction method, the analysis was
carried out using a new method where real direct photons are measured by
their virtual counterparts~\cite{bib:qmp}.
The idea is that all sources of real photons also produce virtual
photons that decay into $e^+e^-$ pairs with small invariant masses. 
The invariant mass distribution of $e^+e^-$ pairs from virtual photons
can be given by Knoll-Wada formula~\cite{bib:knollwada} for each virtual
photon sources, such as direct photon, $\pi^0$ Dalitz and $\eta$ Dalitz.
The ratio of direct photons was obtained using the ratio of photon yield
in different $e^+e^-$ invariant mass region.
Compared to the conventional subtraction measurement, the new 
technique (internal conversion method) improves both the
signal-to-background ratio and the energy resolution at intermediate
$p_\mathrm{T}$. 
An excess of the signal over background has been seen as the red points
of left panel in Fig.~\ref{fig:midpt}, which is consistent with the
result of conventional subtraction method within the error. 

The right panel of Fig.~\ref{fig:midpt} shows the direct photon spectrum 
which is obtained from the excess ratio measured with internal
conversion method and inclusive photon spectrum measured with EMCal.
The spectrum shows the enhancement comparing with the next-to-leading-order
(NLO) pQCD calculation scaled by the Au+Au nuclear overlapping
function~($T_\mathrm{AA}$). 
$T_\mathrm{AA}(b)$ is defined as $T_\mathrm{AA}(b) =
N_\mathrm{coll}(b)/\sigma_\mathrm{NN}$, 
where $N_\mathrm{coll}(b)$ is the average number of binary
nucleon-nucleon collisions at an impact parameter $b$ with an inelastic
cross section~$\sigma_{NN}$.  
A hydrodynamical prediction which computes the thermal photon emission
from a QGP with initial temperatures $\langle T_0 \rangle$ =
360~MeV~\cite{bib:thermal}   
supports the observed enhancement of the spectrum. However, the
reference baseline in those calculations is the $T_\mathrm{AA}$-scaled
NLO pQCD photon spectrum. To confirm the existence of a thermal
enhancement, it is necessary to directly measure the p+p direct photon
spectrum in the $p_\mathrm{T}$ = 1 -- 4~GeV/$c$ range.

\section{Measurement of high-$p_\mathrm{T}$ direct photon}
\label{hard}

\begin{figure}[h]
 \begin{center}
  \begin{minipage}{0.51\linewidth}
   \begin{center}
    \includegraphics*[width=\linewidth]{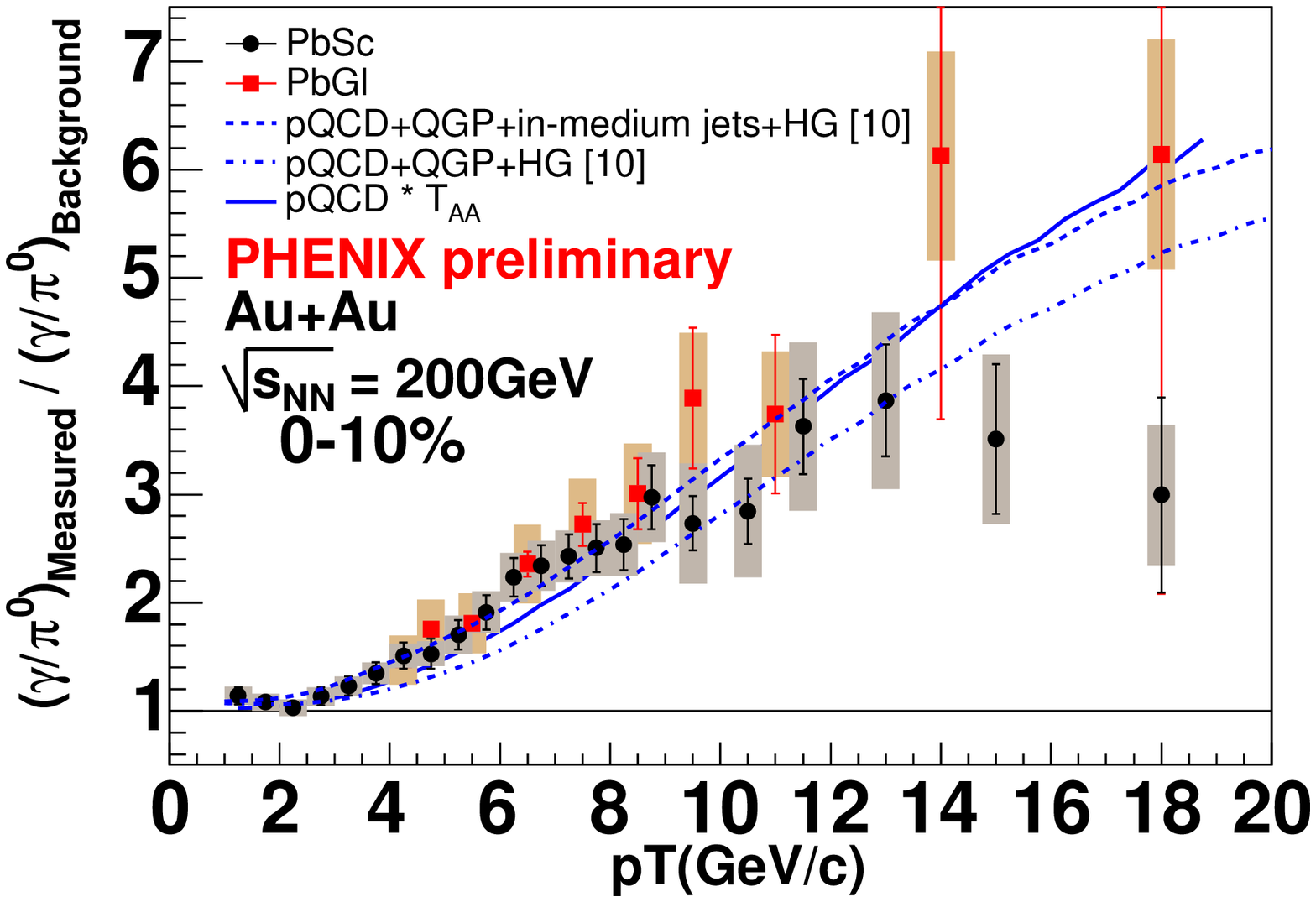}
%    [a]
   \end{center}
  \end{minipage}
%    \hspace{0.05\linewidth}
  \begin{minipage}{0.48\linewidth}
   \begin{center}
    \vspace{-8mm}  % to adjust space between the figure and caption
    \includegraphics*[width=\linewidth]{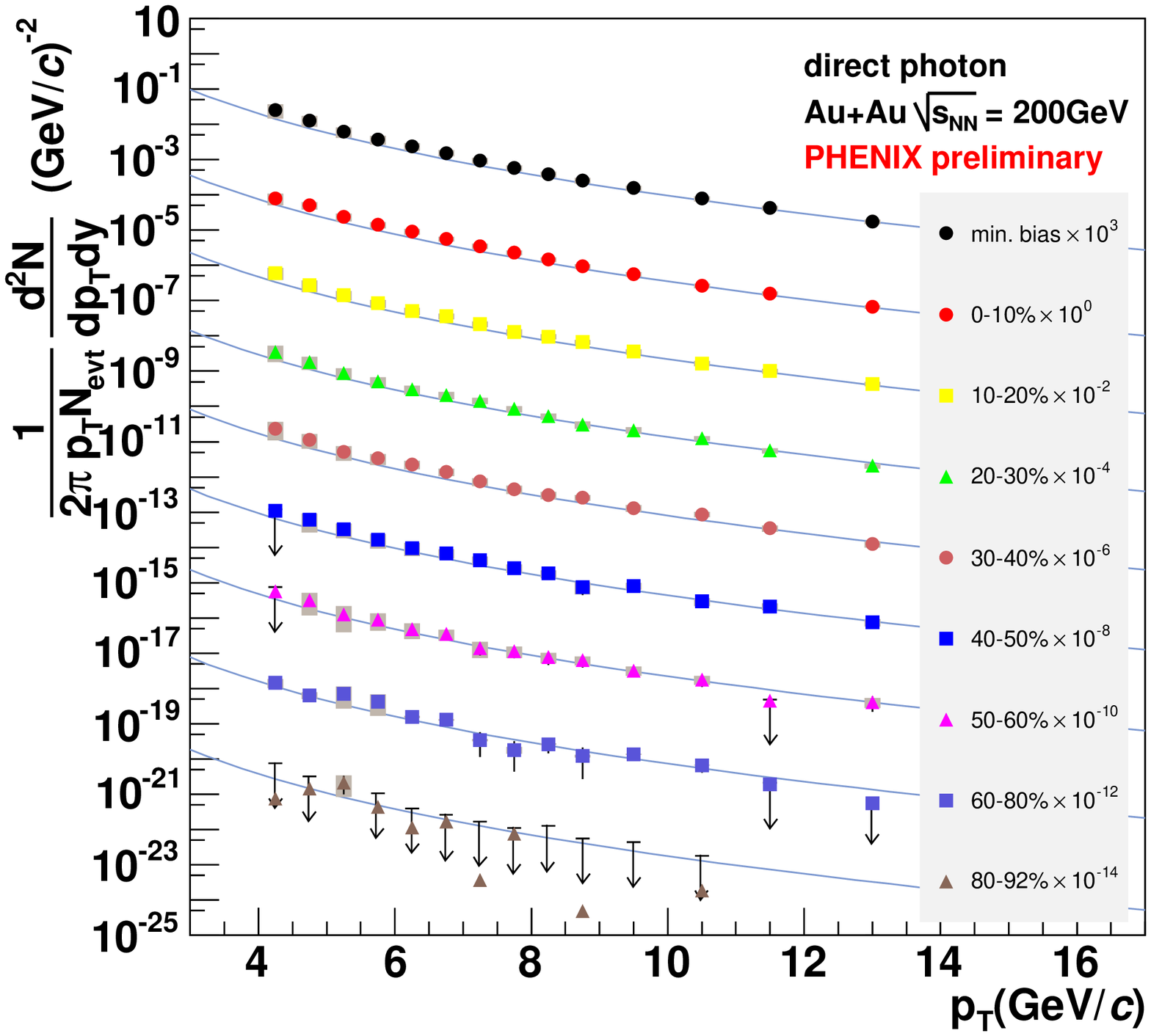}
%    [b]
   \end{center}
  \end{minipage}
 \end{center}
 \caption{Left: Direct photon excess ratio up to high-$p_\mathrm{T}$
 measured with subtraction method. Three types of solid curves are
 theoretical curve: NLO pQCD scaled by $T_\mathrm{AA}$, and a
 calculation which takes into account the jet-quenching of direct
 photons and photons from jet-plasma interaction. Right: The direct
 photon spectra for different centrality classes and minimum bias. The
 solid curves are NLO pQCD calculations scaled by $T_\mathrm{AA}$.}
 \label{fig:highpt}
\end{figure}

In contrast to the measurement of mid-$p_\mathrm{T}$ direct photon, the
strong suppression of neutral hadrons by a jet-quenching effect in heavy
ion collisions~\cite{bib:pi0} allows to extract direct photons
at the momentum region of $p_\mathrm{T} >$ 5~GeV/$c$~\cite{bib:photon}.   
The left panel of Fig.~\ref{fig:highpt} shows the direct photon
excess ratio as a function of $p_\mathrm{T}$ in most central events
~(0-10\%). 
The ratio is in good agreement with a NLO pQCD
calculation~\cite{bib:pqcd} scaled by $T_\mathrm{AA}$ within the error
and theoretical uncertainty. 
This suggests that the initial-hard-scattering probability is not
reduced.   
The direct photon excess ratio measured by PbSc-EMCal slightly deviates
from the NLO pQCD calculation above 14~GeV/$c$.
The deviation may support that the compensation of jet-quenching and
additional photons by a nuclear effect, such as jet-photon conversion,
is breaking up above 14~GeV/$c$.   
Because the direct photon calculated with pQCD consists of the prompt
photon produced from the hard scattering directly and the jet
fragmentation photons coming from hard-scattered partons, the direct
photon yield measured in Au+Au collisions is naively expected to be
suppressed comparing with NLO pQCD calculation scaled by
$T_\mathrm{AA}$.  
The agreement with scaled NLO pQCD calculations might just be a
coincidence caused by mutually counterbalancing effects like energy loss
and Compton like scattering of jet partons~\cite{bib:turbide}. 
In order to conclude that the deviation attributes to the such nuclear
effect, it is important to measure high-$p_\mathrm{T}$ direct photon in
p+p system and to observe the deviation with PbGl-EMCal as well.

The right panel of Fig.~\ref{fig:highpt} shows the fully
corrected direct photon invariant yield as a function of $p_\mathrm{T}$
for each centrality. 
The spectra are combined results of PbSc measurement and PbGl
measurement. 
The yields are in good agreement with $T_\mathrm{AA}$-scaled NLO pQCD 
calculation up to 14~GeV/$c$. 

\section{Summary}
\label{summary}

 The PHENIX experiment has measured direct photons in Au+Au collisions
 at $\sqrt{s_\mathrm{NN}}$ = 200~GeV. 
 The large amount of data taken in RHIC Run4 has made it possible to
 extend $p_\mathrm{T}$ region up to 18~GeV/$c$.
 The result is consistent with the NLO pQCD calculation scaled by the
 nuclear overlap function. 
 The measurement using internal conversion of photons into $e^+e^-$
 shows the enhancement of the yield comparing with NLO pQCD
 calculation. 

% \appendix

% \section{}
% \label{}

\end{document}